\documentclass[10pt,journal,comsoc]{IEEEtran}
\usepackage{url}

\usepackage[noadjust]{cite}
\usepackage{graphicx}
\usepackage{multirow}

\usepackage{subcaption}
\usepackage{xspace}

\usepackage{makecell}

\usepackage{rotating}
\usepackage{lscape}
\usepackage{balance}
\usepackage[utf8]{inputenc}

\usepackage{lipsum}
\usepackage{array, tabularx, multirow, booktabs, diagbox}
\usepackage{adjustbox}
\usepackage{amsmath,amssymb,amsfonts}

\begin{document}
\title{Digital-Twin-Enabled 6G: Vision, Architectural Trends, and Future Directions}

\author{Latif~U.~Khan,~Walid~Saad,~\IEEEmembership{Fellow,~IEEE},~Dusit~Niyato,~\IEEEmembership{Fellow,~IEEE},~Zhu~Han,~\IEEEmembership{Fellow,~IEEE},~and~Choong~Seon~Hong,~\IEEEmembership{Senior~Member,~IEEE}

\IEEEcompsocitemizethanks{
\IEEEcompsocthanksitem This work has been submitted to the IEEE for possible publication. Copyright may be transferred without notice, after which this version
may no longer be accessible.
\IEEEcompsocthanksitem L.~U.~Khan~and~C.~S.~Hong are with the Department of Computer Science \& Engineering, Kyung Hee University, Yongin-si 17104, South Korea.
\IEEEcompsocthanksitem Walid Saad is with the  Wireless@VT, Bradley Department of Electrical and Computer Engineering, Virginia Tech, Blacksburg, VA 24061 USA.
%, and also with the Department of Computer Science and Engineering, Kyung Hee University, Seoul 02447, South Korea. 
\IEEEcompsocthanksitem D. Niyato is with the School of Computer Science and Engineering, Nanyang Technological University, Singapore.
\IEEEcompsocthanksitem Zhu Han is with the Electrical and Computer Engineering Department, University of Houston, Houston, TX 77004 USA, and the Department of Computer Science and Engineering, Kyung Hee University, South Korea.

}}

\markboth{}{}%

\maketitle

% \maketitle

% }

% \thanks{
% 	}
% }

 %The paper headers
\markboth{IEEE Communications Magazine}{}%

% Traditional machine learning is based on the migration of data from massively distributed IoT devices to a centralized cloud for training and thus, poses serious security and privacy concerns.

% \IEEEcompsoctitleabstractindextext{%
% \justify
\begin{abstract} 
Internet of Everything (IoE) applications such as haptics, human-computer interaction, and extended reality, using the sixth-generation ($6$G) of wireless systems have diverse requirements in terms of latency, reliability, data rate, and user-defined performance metrics. Therefore, enabling IoE applications over $6$G requires a new framework that can be used to manage, operate, and optimize the 6G wireless system and its underlying IoE services. Such a new framework for $6$G can be based on digital twins. Digital twins use a virtual representation of the $6$G physical system along with the associated algorithms (e.g., machine learning, optimization), communication technologies (e.g., millimeter-wave and terahertz communication), computing systems (e.g., edge computing and cloud computing), as well as privacy and security-related technologists (e.g., blockchain). First, we present the key design requirements for enabling $6$G through the use of a digital twin. Next, the architectural components and trends such as edge-based twins, cloud-based-twins, and edge-cloud-based twins are presented. Furthermore, we provide a comparative description of various twins. Finally, we outline and recommend guidelines for several future research directions.                      
\end{abstract}

\begin{IEEEkeywords}
Digital twin, $6$G, Internet of Everything, artificial intelligence, distributed machine learning. 
\end{IEEEkeywords}

%In hierarchical federated learning, initially sub-global models are computed at small cell base stations in an iterative manner similar to traditional federated learning. Then, the sub-global models are sent to the macro base station where global model aggregation takes place. Finally, global model updates are sent back to the small cell base stations which further disseminate them to end-devices. The main advantage of hierarchical FL is reuse of the already occupied frequency bands by other users within small cells.

\section{Introduction}
\setlength{\parindent}{0.7cm}The wireless research landscape is rapidly evolving in order to cater for emerging Internet of Everything (IoE) applications such as extended reality (XR), haptics, brain-computer interaction, and flying vehicles \cite{khan20206g}. In order to meet the diverse requirements (e.g., latency, reliability, and user defined metrics) of these IoE applications, the sixth generation ($6$G) of wireless systems  must possess several key properties \cite{6Gvision, kato2020ten, 9145564}: 
\begin{itemize}
\item \textbf{Self-sustaining wireless systems:} $6$G wireless systems will rely on a ubiquitous, intelligent, and seamless connectivity for a massive number of devices to offer various novel IoE services. These IoE services will require true adaptation to the dynamically changing environment and optimization of scarce computation and communication resources. Therefore, to enable IoE services using $6$G, there is a need to propose self-sustaining wireless systems that can operate with minimum possible intervention from end-users. Such self-sustaining systems can jointly perform efficient network functions adaptation and resource optimization, using emerging techniques in the fields of machine learning, optimization theory, game theory, and matching theory, among others.    
\item \textbf{Proactive-online-learning-based wireless systems:} $6$G-based IoE applications must meet highly dynamic and extreme requirements in terms of latency, reliability, data rate, and user-defined performance metrics. To meet these highly dynamic requirements, $6$G will use high frequency bands (e.g. millimeter wave, sub-terahertz, and terahertz), emerging computing technologies (e.g., cloud and edge computing), and security related technologies (e.g., blockchain), among others. Therefore, to successfully enable interaction among these technologies, one cannot rely on classical offline learning systems, but, instead, there is a need for online solutions that can proactively adapt to the 6G system dynamics.  
\end{itemize}\par
\begin{figure*}[!t]
	\centering
	\captionsetup{justification=centering}
	\includegraphics[width=17cm, height=7cm]{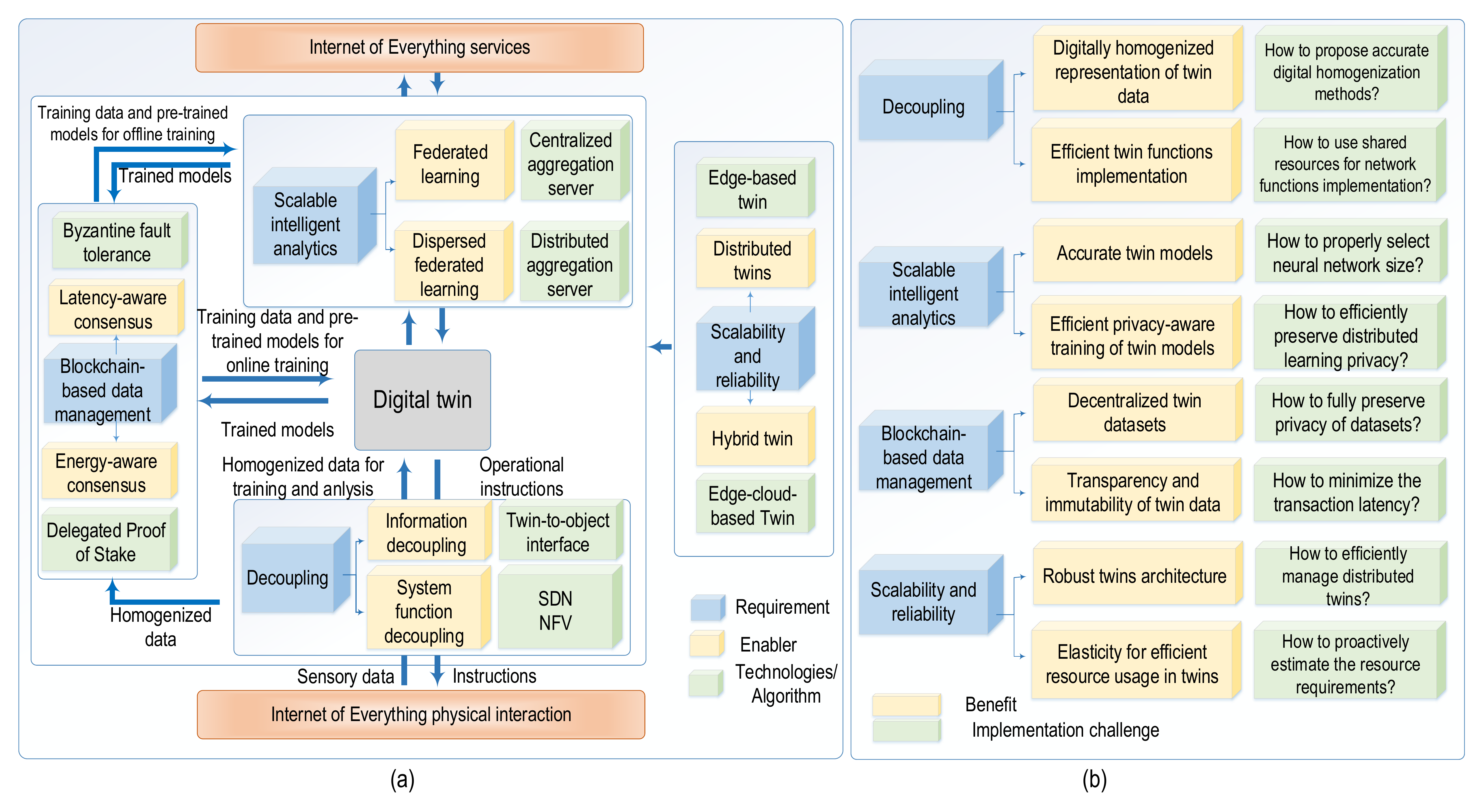}
	\caption{(a) Key design requirements and (b) Benefits for twins-based architecture.}
	\label{fig:requirement}
\end{figure*}\par
To efficiently enable the aforementioned properties of $6$G wireless systems, we can explore the concept of a \emph{digital twin}. A digital twin is a virtual representation of the elements and dynamics of a physical system (e.g., see \cite{yaqoob2020blockchain} and \cite{kritzinger2018digital}). Digital twin uses machine learning, data analytics, and multi-physics simulation to study of the dynamics of a given system. Digital twins can be categorized into: monitoring digital twin, simulation digital twin, and operational digital twin\protect\footnote{Accessed on: $7^{th}$ Oct. 2020,~https://xmpro.com/digital-twins-the-ultimate-guide//}. A monitoring twin supports monitoring the status of a physical system (e.g., autonomous car dashboard), whereas a simulation twin uses various simulation tools and machine learning schemes to provide insights about future states. Meanwhile, an operational twin enables system operators to interact with a cyber-physical system and perform different actions in addition to analysis and system design. Several works considered digital twins for enabling Industry $4.0$ \cite{yaqoob2020blockchain, kritzinger2018digital, 9082917}. In \cite{yaqoob2020blockchain}, the authors devised a taxonomy of blockchain in enabling digital twins. Meanwhile, the authors in \cite{kritzinger2018digital} reviewed digital twins for Industry $4.0$, whereas the work in \cite{9082917} proposed digital twin architecture for Industry $4.0$. In contrast to \cite{yaqoob2020blockchain, kritzinger2018digital, 9082917}, we explore the role of the digital twin in enabling $6$G wireless systems. For the design of $6$G systems, an operational digital twin allows us to enable efficient interaction among various players of $6$G. This is not possible without such an operational twin because other twins (e.g., simulation and monitoring twins) can only enable us to analyze the cyber physical system without controlling the system in real time. As such, our focus will be on the use of an operational digital twin for $6$G. In particular, the digital twin will use a virtual representation of the $6$G physical system along with communication, computing, security, and privacy technologies to enable $6$G-based IoE applications. Meanwhile, to efficiently operate $6$G, the digital twin will use machine learning and optimization algorithms for communication and computing resource optimization. Unless stated otherwise, hereinafter, the term digital twin refers to the operational twin. Our main contributions are as follows.\par
\begin{itemize}
    \item We present the key design requirements to realize the vision of digital-twin-enabled $6$G systems. These requirements are decoupling, scalable intelligent analytics, blockchain-based data management as well as scalability and reliability.
    \item We propose an architecture for digital-twin-enabled $6$G and present its various trends based on deployment fashion of twins. 
    \item Finally, we provide an outlook on future research directions.
\end{itemize}\par
To the best of our knowledge, this is the first work to study digital-twin-enabled $6$G systems. In contrast to existing works \cite{6Gvision,khan20206g, kato2020ten} on $6$G, this paper presents a new perspective regarding the development of intelligent, self-sustaining $6$G wireless systems enabled by digital twin.\par
\section{Key Design Requirements}
In this section, we present the key requirements that are indispensable for creating a digital twin for wireless $6$G systems, as shown in Fig.~\ref{fig:requirement}. These key requirements are decoupling, scalable intelligent analytics, and blockchain-based data management. To fulfill the aforementioned requirements, one must use scalable and reliable architecture and algorithms. A digital twin will use a virtual representation of the system to analyze the system dynamics. For example, in 6G, the digital twin will virtually represent: a) the physical wireless system (e.g., intelligent reflecting surfaces with backhaul links), b) a typical $6$G application physical system (autonomous driving car, Industry $4.0$ plant, and e-healthcare system), and c) specific module (e.g., edge caching module, computational offloading module at the edge). Based on the analysis performed by digital twin for a user requested IoE service, necessary actions will be performed to serve the $6$G service requesting user. To carry out analysis and control actions, there is a need to efficiently deploy digital twin models with other technologies to enable easier network management (more details about architecture will be given in Section~\ref{architecture}). To train digital twin models, one can use the data generated by IoE devices. In addition to the IoE devices data, the blockchain network will store pre-trained twin models. These pre-trained models can be employed for training other twin models using newly received data for better performance. This way of updating the twin models will improve the overall performance over time by using newly added data. Next, we discuss the requirements for enabling digital twin for $6$G in detail.\par
\begin{table*}
	\caption {Comparison of edge-based, cloud-based, and edge-cloud-based twins.} 
	\label{tab:edgecomparison} 
	\centering
	\begin{tabular}{p{1.5cm}p{9cm}p{2.4cm}p{1.5cm}p{1.5cm}}
		\toprule
		&\textbf{Description}&\textbf{Edge-based twin} & \textbf{Cloud-based twin} & \textbf{Edge-cloud-based twin} \\
		\hline
		\textbf{Scalability} & Scalability refers to fulfilling latency requirements for massive number of $6$G devices. Furthermore, the addition of new nodes should not significantly degrade the system performance in terms of latency. & High & Lowest & Low \\ \midrule
		\textbf{Latency} & This metric represents the overall delay in providing $6$G services. & Low & High & Medium \\ \midrule
		\textbf{Geo-distribution} & This metric tells us about the geographical distribution of twin objects for enabling a $6$G service. & Distributed & Centralized & Hybrid \\\midrule
		\textbf{Elasticity} & This metric refers to on-demand dynamic resource allocation for digital twins operation in an elastic way in response to highly dynamic requirements. & High & Low & High \\ \midrule
		
		\textbf{Context-awareness} & Context-awareness is the function that deals with the knowledge about the end-devices location and network traffic. & High & Low & Medium \\\midrule
		\textbf{Mobility support} & Mobility support deals with the ability of digital twins to seamlessly serve mobile end-devices. & High & Low & Medium \\ \midrule
		\textbf{Robustness} & Robustness refers to seamless operation of digital-twin-enabled $6$G application in case of failure one of the twin objects. & Highest (for multiple edge-based twins) & Lowest & Medium \\
		\bottomrule
	\end{tabular}
\end{table*}\par

\subsection{Decoupling}
The transformation of a physical system into a digital twin is primarily based on \emph{decoupling}. Decoupling in the context of digital twins can be of two types: information decoupling and system functions decoupling, as shown in Fig.~\ref{fig:requirement}a. Information decoupling allows the transformation of physical system information (i.e., system state) into a homogenized digital representation that will offer generality and ease the implementation of digital twins. Moreover, homogenized digital representation helps digital twins to easily realize the dynamically changing states of a physical interaction space. A $6$G physical interaction space consists of base stations (BSs), intelligent reflecting surfaces, smart devices/sensors, and edge/cloud servers. Examples of $6$G physical interaction space information inlcude industrial process control sensory data and holographic images, haptics sensors, $6$G spectrum usage data, edge servers resource management data, among others. On the other hand, in order enable a flexible operation of a digital twin-based wireless system, it is necessary to effectively decouple the system functions (i.e., mobility management, resource allocation, edge caching, etc.) from hardware to software. The decoupling of system functions will allow us to make the digital twin-based system operate efficiently and adaptively as per the network dynamics (i.e., network adaptability to enable self-sustainable $6$G networks). To enable digital twin for $6$G with functions decoupling, software-defined networking (SDN) and network function virtualization (NFV) can be the promising candidates. SDN offers separation between control plane and data plane, whereas NFV offers system cost-efficient implementation of various network functions using virtual machines running on generic hardware. Although SDN and NFV are key enablers of network slicing, digital-twins-enabled $6$G will be different than the classical network slicing. In contrast to network slicing, digital twin-enabled $6$G will use a digital representation of the physical system to enable $6$G services. The digital representation with machine learning will enable us to proactively analyze and model various system functions. These trained models will be stored in a blockchain network for further use (will be discussed in more detail in Section~\ref{Blockchain-Based Data Management}). Overall, the digital twin-based $6$G system can be seen as a complex concept that performs offline analysis (e.g., proactive analytics, such as data analytics and pre-training of twin models, using stored data on a blockchain) and real-time control. However, network slicing will enable real-time resource management in response to the end-users requests. Therefore, we can say that a digital twin-based $6$G system will use network slicing in addition to other technologies (e.g., data decoupling, interfacing, blockchain, proactive analytics, optimization) for efficient control. \par  

\subsection{Scalable Intelligent Analytics}
\label{Scalable Intelligent Analytics}
$6$G must be able to sustain heterogeneous system requirements, network structures, and hardware architectures. Therefore, digital-twin-enabled $6$G wireless systems should be based primarily on effective machine learning schemes for large datasets \cite{kato2020ten}. However, training twin models for large datasets faces many challenges, such as the need to deal with a complex machine learning model of large size and the high computing power needed for training. For instance, shallow neural networks can have better scalability in terms of computing power for training large datasets, but their performance will degrade for highly dynamic scenarios (e.g., for mobility management, resource allocation, and edge caching). Although one can consider a deep learning-based twin model for highly dynamic scenarios with large datasets, training of a deep learning-based twin model at a centralized location might not be feasible due to a high training time. Moreover, the inference will also be slow for large datasets. To address these challenges, we can use distributed deep learning-based twin models. In distributed deep learning models, multiple models can be trained at various locations to reduce the model training time. After the computation of multiple twin models at different locations, all the trained twin models are combined at a centralized location. This process continues in an iterative manner until convergence. Generally, twin model computing time and communication time have an inverse relationship: When the number of distributed machines is increased, the model computing time decreases while the communication time increases. However, there is some saturation point, beyond which the training cost (i.e., the sum of computation time and communication time) shows no significant change for an increase in the number of machines \cite{wen2017terngrad}. Therefore, we must define new scalable machine learning schemes to overcome this limitation. Federated learning with sparsification can be a promising solution to enable scalable, distributed machine learning-based twin model. Sparsification-enabled federated learning sends only the important values of a full gradient, and thus further reduces the communication overhead. Although federated learning offers scalable machine learning, it has robustness issues due to its dependency on a single centralized server for global model computation. The centralized aggregation server might fails due to an attack or physical damage. To address this limitation, we can use dispersed federated learning based on distributed aggregations \cite{khan2020dispersed}.\par
\subsection{Blockchain-Based Data Management}
\label{Blockchain-Based Data Management}
Digital-twin-enabled $6$G will be based mostly on decentralized network architectures using machine learning to offer extremely low latency services. To manage the decentralized datasets in a transparent and immutable manner, blockchain is a promising candidate. This is because blockchain allows us to store data that cannot be altered without collusion of the network majority and a retroactive change of all subsequent blocks. Moreover, blockchain can offer a seamless transfer of data without losing data integrity. Specifically, blockchain can be used to store the twins pre-trained machine learning models for various scenarios which can further use performance improvement in terms of learning accuracy and reduced training time. Additionally, it can store the data required for the training of digital twin machine learning models. For example, a blockchain can be used to store pre-trained models for medical imaging applications developed by different hospitals and healthcare centers. Moreover, blockchain can store additional datasets from new clinics as it can be considered as a trusted data management platform. \par 
To run a blockchain consensus algorithm, edge servers can be used as miners. Although blockchain can offer several benefits, it has a few challenges. These challenges are scalability, the high latency associated with a blockchain consensus algorithm, high-energy consumption, and privacy concerns \cite{yaqoob2020blockchain}. Generally, blockchain has a slower transaction speed with an increase in the number of nodes. Therefore, we must resolve this bottleneck of the blockchain to enable it for digital twins-enabled $6$G. The consensus algorithm used in blockchain adds a significant amount of delay and consumes energy before reaching a consensus. Furthermore, it has privacy concerns because of its distributed design. Every node in a blockchain has access to the blockchain transaction data itself, and thus privacy leakage can occur in the presence of a malicious user. To efficiently operate digital-twin-enabled $6$G, we must propose blockchain consensus algorithms that will offer low-latency (e.g., Byzantine fault tolerance) and low energy (e.g., delegated proof of stake). \par             
\subsection{Scalability and Reliability}
The expected massive number of devices in $6$G motivates us to design a scalable and reliable architecture based on digital twins. Specifically, digital twins will suffer from a scalability and reliability issue, when it comes to implementing massive ultra-reliable low latency communication (mURLLC) services \cite{6Gvision}. Although cloud-based twin implementation can offer a low design complexity pertaining to management and design, it will suffer from high latency issues for networks with a large number of devices. To address this issue, one can use a distributed twin architecture (as detailed in Section~\ref{architecture}). Distributed architectures (i.e., edge-based twins) reduce latency and increase scalability. However, distributed digital twins will suffer from higher management complexity compared to centralized digital twins. A centralized digital twin may have more computational power and storage than the distributed one, but it has high latency. Therefore, we can use a hybrid approach by combining the features of both centralized and distributed digital twins to offer a tradeoff between computational power, storage capacity, and latency. For instance, consider collaborative caching for extended reality. One can deploy edge-based twin objects using deep reinforcement learning at the network edge and cloud-based twin object at a cloud. Edge-based twin objects will update themselves using their own data and send their learning updates to the cloud-based twin. Finally, cloud-based twin object can share the updated learning updates with all edge-based twin objects. \par

\section{Architecture of Digital-Twin-Enabled 6G}
\label{architecture}
\begin{figure*}[!t]
	\centering
	\captionsetup{justification=centering}
	\includegraphics[width=16cm, height=11cm]{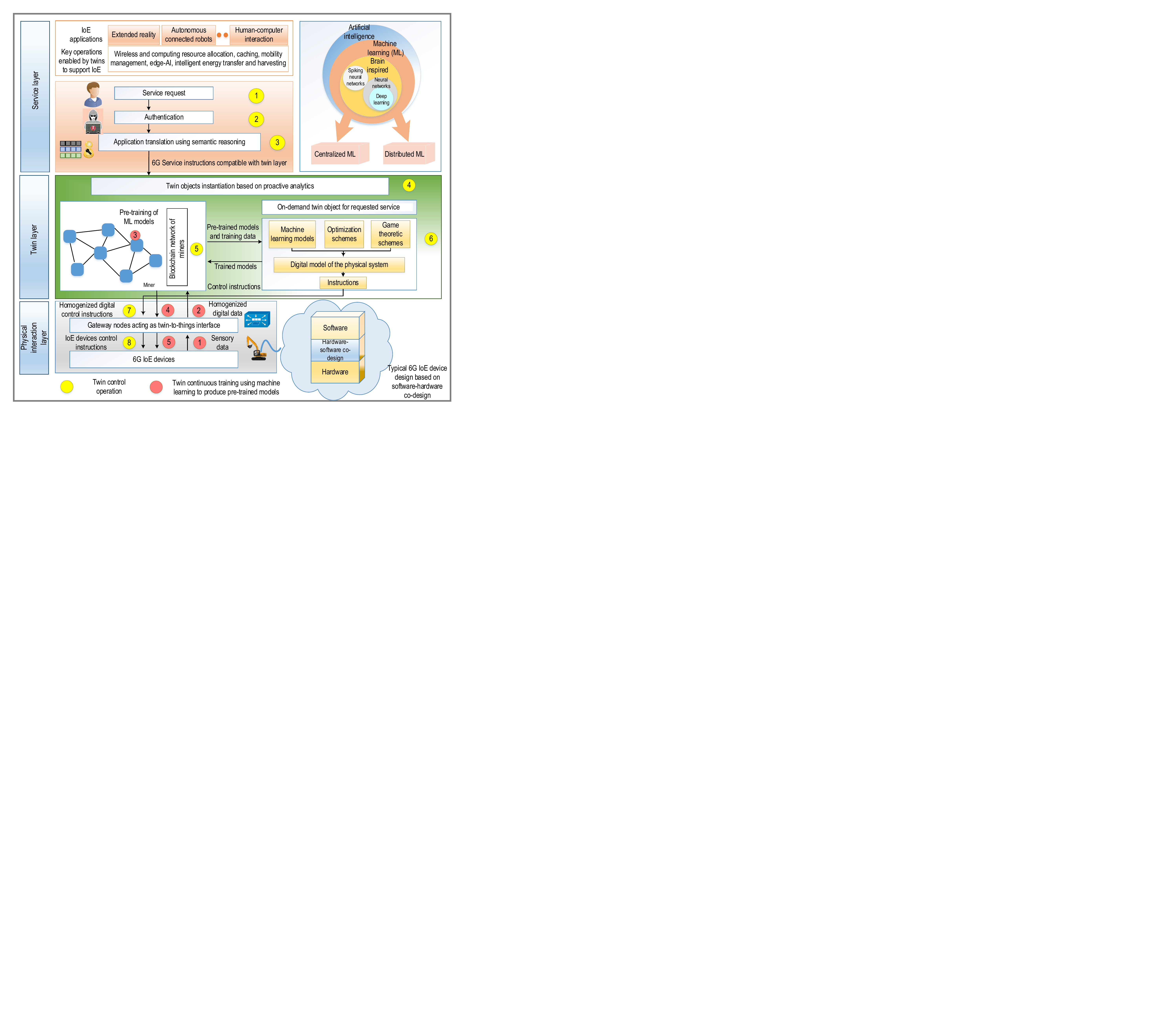}
	\caption{Digital twin sequence diagram.}
	\label{fig:Twin_cycle}
\end{figure*}

We propose the notion of virtual twin objects-based architecture for digital-twin-enabled $6$G to enable various IoE applications, whose overview is given in Fig.~\ref{fig:Twin_cycle}. The twin objects interact with IoE devices and other twins to enable $6$G services. We next present a detailed discussion on digital-twin-based $6$G architecture components and its deployment possibilities.\par
\subsection{Twin Objects}
To efficiently deploy digital-twin-enabled $6$G wireless systems, we can use the notion of virtual twin objects. Twin objects will be responsible for performing optimization, training of a machine learning model, and control to enable a given $6$G service. A typical $6$G service can be deployed either using a single or multiple twin objects. These twin objects can be created and terminated dynamically to enable $6$G services based on requests \cite{6525602}. Twin object can be implemented using transient-based virtual machines (TVM). TVM-based twin objects will offer us the proactive customization of resources and post-use cleanup. The proactive customization of resources can be enabled by proactive intelligent analytics based on emerging machine learning schemes \cite{kato2020ten}. Later, the resources assigned to the TVM should be released after executing the service. It must be noted that the TVM for a particular service should be isolated from the TVM of another service. Furthermore, the TVM must be separated from the host node software. The host node software must interact with the TVM seamlessly and ubiquitously to enable us with the efficient implementation of multiple TVMs. \par   
\begin{figure}[!t]
	\centering
	\captionsetup{justification=centering}
	\includegraphics[width=8cm, height=5cm]{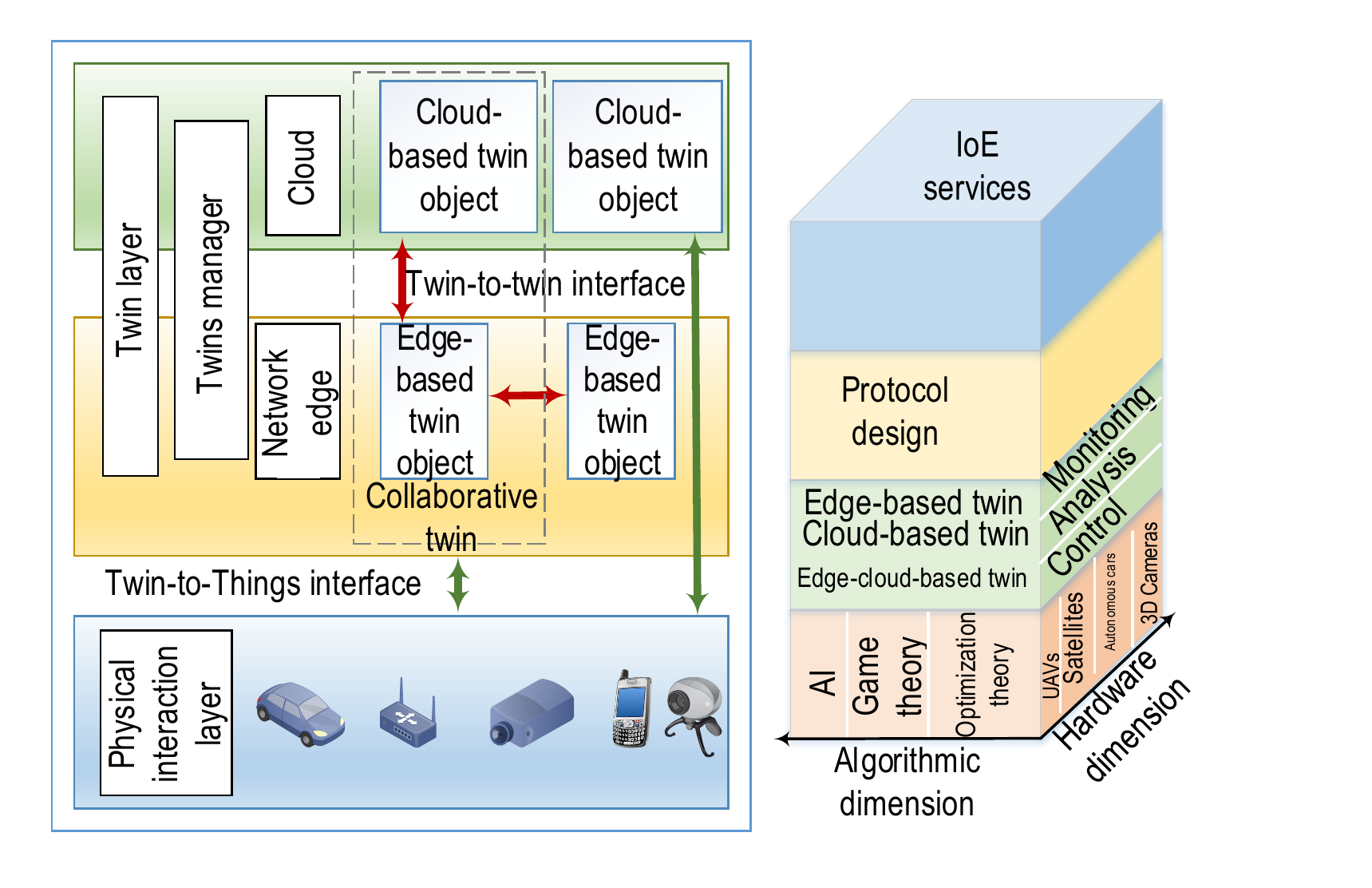}
	\caption{Twin objects deployment trends.}
	\label{fig:trends}
\end{figure}
\par
\subsection{Twin Object Deployment Trends}
Twin objects can be deployed either at end-devices, edge, or cloud. Depending on the deployment of twin objects, we can divide the architecture of digital twin into three categories: \emph{edge-based digital twin, cloud-based digital twin, and edge-cloud-based collaborative digital twin}, as shown in Fig.~\ref{fig:trends}. Edge-based twin objects are more suitable for $6$G applications with strict latency constraints (e.g., massive URLLC), whereas cloud-based twin objects can be used for delay tolerant, high computational power applications. Meanwhile, edge-cloud-based twins can use both edge and cloud resources to offer a tradeoff between latency and computational power. Edge-based twins are characterized by lower computational power and lower storage than cloud-based twins. Edge-cloud-based twins exploit the advantages of both cloud-based twins (i.e., high computational power) and edge-based twins (i.e., instant analytics). For instance, consider a cooperative intelligence-enabled transportation system using edge-cloud-based twins. For designing traffic congestion control in $6$G, we can employ twins that are installed at the cloud \cite{shengdong2019intelligent}, whereas for reporting accidents between autonomous cars, we can use edge-based twins due to the instant reporting requirement of such events \cite{khan2020edge}. Comparison of twins for various performance metrics is given in Table~\ref{tab:edgecomparison}. Considering scalability, an edge-based twin will have the highest value due to the fact that it has the lowest latency compared to both cloud and edge-cloud-based twins. We can add more end-devices in case of edge-based twins until their maximum serving limit without significantly increasing the latency. However, a cloud-based twin will suffer from low scalability due to increase in latency when the number of devices increases. Therefore, depending on the $6$G application nature, we should deploy the twin object at an appropriate location. Furthermore, elasticity for both edge and edge-cloud-based-twins is much higher than for a cloud-based twin because of low context-awareness (i.e., less information about the network edge dynamics). In terms of mobility support, edge-based twins and edge-cloud-based twins have higher support than a cloud. Also, edge-based twins and edge-cloud-based twins have more information about end-user network dynamics, and thus can easily handle mobility than the remote cloud. \par
Various interfaces such as twins-to-things, twin-to-twin, and twin-to-service interfaces must be proposed for seamless, isolation-based, complex interaction in a scalable and reliable manner. Using a twin interface, we can add and remove twins designated for specific applications without affecting the other twins' operation \cite{sha2020survey}. Furthermore, twin-to-object interfaces allow us to efficiently decouple the IoE devices from the twin layer, thus proving us with easier management. A twin-to-twin interface enables the communication between various twin objects to implement a distributed system (e.g., federated learning-based vehicular edge computing \cite{8964354}). For instance, to implement distributed machine learning-based systems, various twin objects can be used to train machine learning models at network edges. Next, to get an ensemble machine learning model, a twin-to-twin interface will be used. Furthermore, a twin-to-twin interface will be used for communication among various twins at different levels (i.e., cloud and edge levels).\\

\subsection{Digital Twin Operational Steps}
Mainly, we can divide the twin operation into two types: training and operation as shown in Fig.~\ref{fig:Twin_cycle}. For training, one can use distributed machine learning (step $1$). Next, the local learning models are sent to the twin layer for aggregation at the blockchain miner (steps $2$ and $3$). After the computation of the global model, the global model updates are sent back to the IoE devices for updating their local learning models (steps $4$ and $5$). This iterative process of learning can take place either in a synchronous or asynchronous fashion. In an asynchronous fashion, a device will send its local learning model only when getting a connection to miners, whereas the devices, must send their local learning models within a predefined time to the miner for global aggregation in case of synchronous fashion. Therefore, we must appropriately select the fashion of aggregation depending on connectivity conditions. On the other hand, there are some scenarios such as autonomous driving cars in which devices can generate up to $4,000$ gigaoctet of data every day, which must be considered in the training of the considered learning model. Although one can use centralized machine learning for such scenarios, it has the downside of using more communication resources to transfer the end-devices data to a centralized server. To address this challenge, we should use federated learning to continuously update the global model for better performance.\par  
Now we explain the operation of digital twin in response to the end-users request for $6$G service (XR is used an example here to provide a concrete understanding of the digital twin operation). For instance, consider a $6$G system in which an XR device requests a service from the BS (step $1$). In response to the user request, first of all, authentication takes place (step $2$). After validation of the end-user XR request, the XR request must be translated using semantic reasoning techniques, into a form understandable by the twin objects (step $3$). Next, we must instantiate twin objects at the BS based on TVM and associate them with blockchain miners (step $4$). The miners will store and run a blockchain consensus algorithm to enable trustworthy sharing of data for twins' operations (step $5$). Moreover, miners will store the data required for twin proactive analytics. Additionally, miner will store the pre-trained models for XR service. Upon the end-user request for services, the on-demand twin objects can use these pre-trained models (i.e., either use them directly for instant XR operation and perform further training for future use). The instantiated twin object will serve the XR requesting end-user by enabling efficient communication and computing resource management with other controls (steps $6$, $7$, and $8$). \par
\section{Future Directions}
\subsection{Isolation Between Twins-Based Services}
To allow $6$G to use twin objects for various IoE applications, efficient utilization of network resources is necessary. To guarantee the performance of a twin-based IoE service without affecting the performance of other twin-based services, there is a need to explicitly allocate resources (i.e., computation and communication resources) for various twin-based services. One way can be to use a dedicated allocation of resources to various twin-based services. However, dedicated allocation of resources (e.g., edge server) to a digital twin-based service will result in an inefficient use of resources. Therefore, we must propose novel optimization schemes for twin objects that will enable the efficient use of resources by sharing them among many twin-based services while fulfilling their isolation requirements. \par
%{\em How do we effectively enable a twin-based service without affecting the operation of other twin-based services using shared resources to enable a cost-efficient $6$G system?} To allow digital twins to use twin objects for various applications, efficient utilization of network resources is necessary. To guarantee the performance of a twin without affecting the performance of other twins, there is a need to explicitly allocate resources (i.e., computation and communication resources) for various twins. One way can be dedicated allocation of resources to various twins. However, dedicated allocation of resources (e.g., edge server) to a digital twin-based service will result in an inefficient use of resources. Therefore, we must propose novel schemes that enable the efficient use of resources by sharing them among many twin objects while fulfilling the isolation requirements of twin-based services. \par

\subsection{Mobility Management for Edge-Based Twins}
An end-user must be served seamlessly by the digital-twin-enabled $6$G system during the service period. A mobile device associated with a twin object might suffer from service interruption due to moving outside the coverage of the access point/BS associated with the twin. One way is to enable the end-user with the service via communication with its twin object/s using a backhaul link. But, this approach will suffer from the issues of slight interruption of the service and high latency. To cope with these issues, one can migrate the service to the newly associated twin objects. However, migrating the service to new twin object/s will be based on the prediction of user mobility to proactively find the new twin objects for service migration. Therefore, we must propose effective prediction schemes to find the locations of newly twin objects for service migration.\par
\subsection{Digital Twin Forensics}
A typical digital-twin-enabled $6$G system will have a variety of players (e.g., end-devices, TVM-based twin-objects, communication interfaces). Therefore, it will be vulnerable to various security threats. To enable the successful operation of the twin-based system, there is a need to propose effective forensic techniques to investigate these security attacks. Based on the analysis of attacks, there is a need to develop new security mechanisms. The main challenges that will be involved in digital-twin-enabled $6$G forensics are attacks evidence identification, evidence acquisition and preservation, and evidence presentation. Evidence identification is the first and difficult to find out due to the proliferation of devices and data, device mobility, heterogeneity in devices, and their software. Other than evidence identification, evidence acquisition, and preservation pose challenges to implement forensics techniques. Acquiring evidence from the devices suffers from data encryption and hardware/software heterogeneity. Next, preserving the evidence for investigation in cases of limited memory devices. One possible solution to preserving evidence is to save it in another location (e.g., edge computing servers). The final step in forensics is how to present the data to enable efficient forensics analysis.           

\section{Conclusions}
In this paper, we have presented a vision of digital-twin-enabled $6$G.We have proposed a digital twin-based architecture for $6$G. Furthermore, we have provided an outlook on future research. We have concluded that a digital twin will serve as a key enabler of $6$G services. Our proposed edge-based digital twins will offer key features scalability and reliability by using distributed deployment. We have also provided a roadmap for future research to truly realize the vision of a digital twin for $6$G.

\bibliographystyle{IEEEtran}
\bibliography{Database}

\begin{IEEEbiography}[{\includegraphics[width=1in,height=1.25in,clip,keepaspectratio]{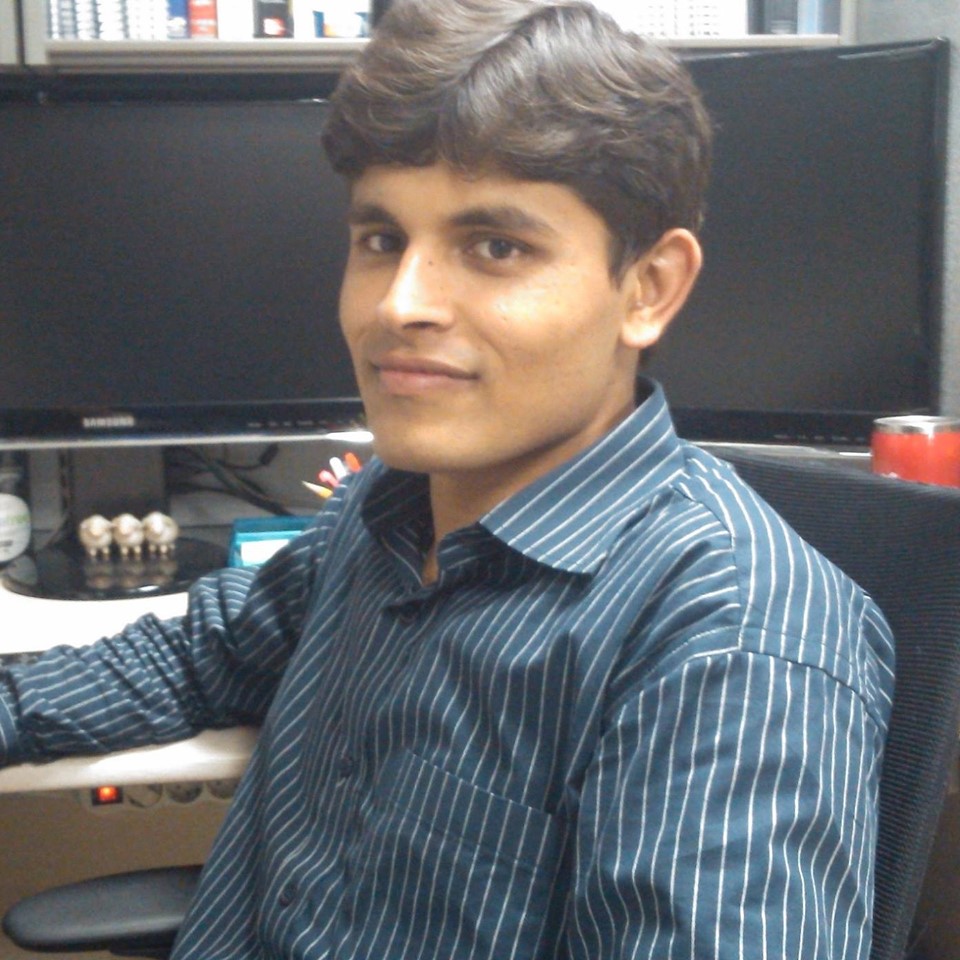}}]{Latif U. Khan} is currently pursuing his Ph.D. degree in Computer Engineering at Kyung Hee University (KHU), South Korea. He received his MS (Electrical Engineering) degree with distinction from University of Engineering and Technology, Peshawar, Pakistan in 2017. His research interests include analytical techniques of optimization and game theory to edge computing and end-to-end network slicing. 
\end{IEEEbiography}
\begin{IEEEbiography}[{\includegraphics[width=1in,height=1.25in,clip,keepaspectratio]{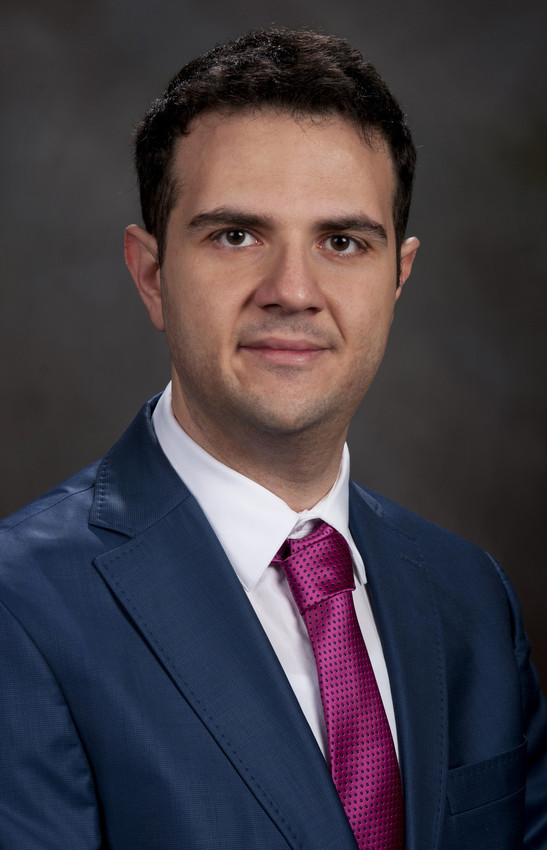}}]{Walid Saad } (S’07, M’10, SM’15, F’19) received his Ph.D. degree from the University of Oslo in 2010. Currently, he is a professor in the Department of Electrical and Computer Engineering at Virginia Tech. His research interests include wireless networks, machine learning, game theory, cybersecurity, unmanned aerial vehicles, and cyber-physical systems. He is the author/co-author of ten conference best paper awards and of the 2015 IEEE ComSoc Fred W. Ellersick Prize. \end{IEEEbiography}

% .

\begin{IEEEbiography}[{\includegraphics[width=1in,height=1.25in,clip,keepaspectratio]{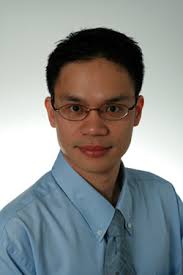}}]{Dusit Niyato} (M’09, SM’15, F’17) is currently a professor in the School of Computer Science and Engineering, Nanyang Technological University. He received his B.Eng. from King Mongkuts Institute of Technology Ladkrabang, Thailand, in 1999 and his Ph.D. in electrical and computer engineering from the University of Manitoba, Canada, in 2008. His research interests are in the area of energy harvesting for wireless communication, Internet of Things (IoT), and sensor networks.

\end{IEEEbiography}

\begin{IEEEbiography}[{\includegraphics[width=1in,height=1.25in,clip,keepaspectratio]{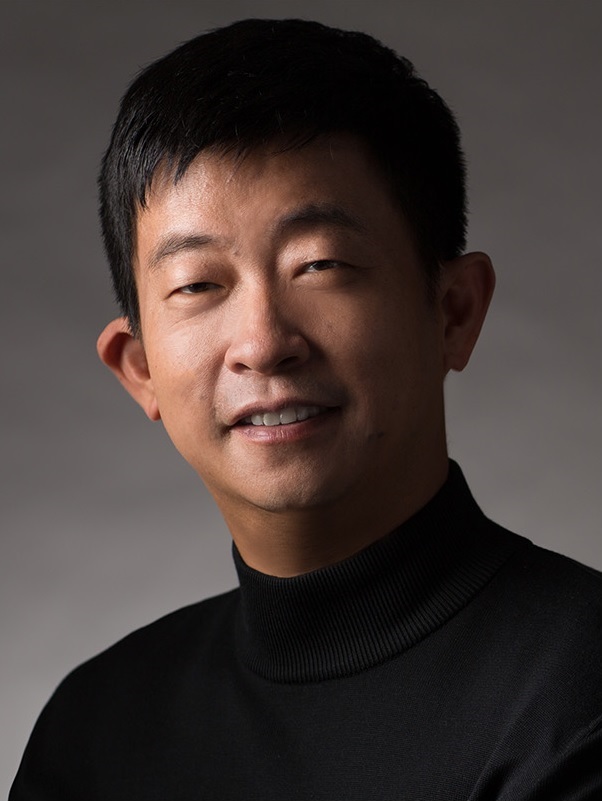}}]{Zhu Han}(S’01, M’04, SM’09, F’14) received his Ph.D. degree in electrical and computer engineering from the University of Maryland, College Park. Currently, he is a professor in the Electrical and Computer Engineering Department as well as in the Computer Science Department at the University of Houston, Texas. Dr. Han is an AAAS fellow since 2019. Dr. Han is 1\% highly cited researcher since 2017 according to Web of Science, and winner of 2021 IEEE Kiyo Tomiyasu Award.

\end{IEEEbiography}

\begin{IEEEbiography}[{\includegraphics[width=1in,height=1.25in,clip,keepaspectratio]{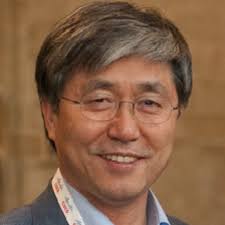}}]{Choong Seon Hong} (S’95-M’97-SM’11) is working as a professor with the Department of Computer Science and Engineering, Kyung Hee University. His research interests include future Internet, ad hoc networks, network management, and network security. He was an Associate Editor of the IEEE Transactions on Network and Service Management, Journal of Communications and Networks and an Associate Technical Editor of the IEEE Communications Magazine.
\end{IEEEbiography}

\end{document}